\documentclass[pre,showpacs,floatfix]{revtex4}
\usepackage{amssymb}
\usepackage{amsmath}
\usepackage{graphicx}
\usepackage{color}

\begin{document}

\title{Nonlinear{\ localized flatband} modes with spin-orbit coupling}
\author{G. Gligori\'c, A. Maluckov, Lj. Had\v zievski}
\affiliation{Vin\v{c}a Institute of Nuclear Sciences, University
of Belgrade, Serbia}
\author{Sergej Flach}
\affiliation{Center for Theoretical Physics of Complex Systems, Institute for Basic
Science, Daejeon, South Korea \\
and \\
New Zealand Institute for Advanced Study, Centre for Theoretical Chemistry
\& Physics, Massey University, Auckland, New Zealand }
\author{Boris A. Malomed}
\affiliation{Department of Physical Electronics, School of Electrical Engineering,
Faculty of Engineering, Tel Aviv University, Tel Aviv 69978, Israel\\
and\\
Laboratory of Nonlinear-Optical Informatics, ITMO University, St. Petersburg
197101, Russia}
\date{\today }

\begin{abstract}
We report the coexistence and properties of stable compact localized states
(CLSs) and discrete solitons (DSs) for nonlinear spinor waves on a flatband
network with spin-orbit coupling (SOC). The system can be implemented by
means of a binary Bose-Einstein condensate loaded in the corresponding
optical lattice. In the linear limit, the SOC opens a minigap between flat
and dispersive bands in the system's bandgap structure, and preserves the
existence of CLSs at the flatband frequency, simultaneously lowering their
symmetry. Adding onsite cubic nonlinearity, the CLSs persist and remain
available in an exact analytical form, with frequencies which are smoothly
tuned into the minigap. Inside of the minigap, the CLS and DS families are
stable in narrow areas adjacent to the FB. Deep inside the semi-infinite
gap, both the CLSs and DSs are stable too.
\end{abstract}

\maketitle

\section{Introduction}

Wave dynamics can be tailored by symmetries and topologies
imprinted by dint of underlying periodic potentials. In turn, the
symmetries and topologies of the periodic potentials can be probed
by excitations in the system into which the potential is embedded.
In particular, flatband (FB) lattices, existing due to specific
local symmetries, provide the framework supporting completely
dispersionless bands in the system's spectrum \cite{richter15}. FB
lattices have been realized in photonic waveguide arrays
\cite{guzman-silva14}, exciton-polariton condensates
\cite{masumoto12}, and atomic Bose-Einstein condensates (BECs)
\cite{taie15}.

FB lattices are characterized by the existence of compact
localized states (CLSs), which, being FB eigenstates, have nonzero
amplitudes only on a finite number of sites \cite{richter15}. The
CLSs are natural states for the consideration of their perturbed
evolution. They feature different local symmetry and topology
properties, and can be classified according to the number $U$ of unit
cells which they occupy \cite{flach14}. Perturbations may
hybridize CLSs with dispersive states through a spatially local
resonant scenario \cite{flach14}, similar to Fano resonances
\cite{miroshnichenko10}. The CLS existence has been experimentally
probed in the same settings where
FB lattices may be realized, as mentioned above: waveguiding arrays \cite%
{vicencio15,mukherjee15,weimann16}, exciton-polariton condensates \cite%
{baboux16}, and atomic BECs \cite{taie15}. The impact of various
perturbations, such as disorder \cite{flach14,leykam16}, correlated
potentials \cite{bodyfelt14,danieli15}, and external magnetic and electric
fields \cite{khomeriki16}, on FB lattices and the corresponding CLSs was
studied too.

A particularly complex situation arises in the case of much less
studied nonlinear perturbations, which can preserve or destroy
CLSs, and detune their frequency
\cite{vicencio13,leykam13,johannson15,lopez-gonzalez16}. Here we
study the existence of nonlinear localized modes in a pseudospinor
(two-component) diamond chain, whose components are linearly mixed
due to spin-orbit-coupling (SOC). The system can be implemented
using a binary Bose-Einstein condensate (BEC) trapped in an
optically imprinted potential emulating, e.g., the "diamond chain"
\cite{flach14}. The two components represent different atomic
states, and the SOC interaction between them can be induced by
means of a recently elaborated technique, making use of properly
applied external magnetic and optical fields \cite{SOC}. The
possibility to model these settings by discrete dynamics in a deep
optical-lattice potential was demonstrated, in a general form, in
Refs. \cite{discreteso,nashsoc}. We consider two types of
nonlinearities produced by interactions between atoms in the BEC,
\textit{viz}., intra- and inter-component ones. The main objective
of the analysis is to analyze the impact of the SOC on the linear
and nonlinear CLS modes, as well as on exponentially localized
discrete solitons. We demonstrate the possibility to create
diverse stable localized modes at and close to the FB frequency,
and inside gaps opened by the SOC.

In a previous work \cite{nashsoc}, we studied the effect of the SOC on the
dynamics of discrete solitons in a binary BEC trapped in a deep
one-dimensional (1D) optical lattice. Among new findings related to the SOC
were the tunability of the transition between different types of localized
complexes, provided by the SOC strength, and the opening of a minigap in the
spectrum induced by the SOC. Inside the minigap, miscible stable on-site
soliton complexes were found \cite{nashsoc}. In the opposite,
quasi-continuum limit, one- and two-dimensional discrete solitons supported
by the SOC were studied too \cite{discreteso}.

The paper is structured as follows. The model is introduced in Section II.
Following a brief recapitulation of the spectral properties of the
single-component linear quasi-1D diamond-chain lattice, the two-component
system is considered. It is shown that the SOC opens gaps between the FBs
and DBs in the spectrum. In Section III, exact solutions for CLS modes are
constructed in the linear system with the SOC terms. Effects of the SOC on
nonlinear CLS modes, and a possibility to create other types of the
localized ones, in gaps between the FB and DB is considered in Section IV.
In particular, the nonlinear CLSs are found in an exact analytical form too.
In that Section, localized modes in the semi-infinite gap (SIG) are briefly
considered too. The paper is concluded by Section V.

\section{Model equations}

\subsection{The single-component model}

\begin{figure}[h]
\centering
\includegraphics[width=12cm]{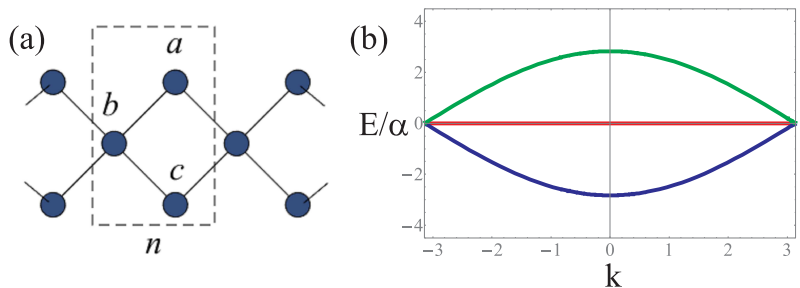}
\caption{(a) The schematic of the diamond chain corresponding to
Eq. (\ref{first}). Circles and solid lines designate lattice
sites, and hoppings, respectively. The dashed rectangle defines
the unit cell, consisting of a (upper), b (middle) and c (bottom)
sites. (b) The dispersion relation for the linear case $\beta =0$
(see details in the text).} \label{fig:diamond}
\end{figure}

We consider the one-dimensional "diamond-chain" lattice shown in
Fig. \ref{fig:diamond}(a). Its bandgap structure, shown in Fig.
\ref{fig:diamond}(b), consists of two DBs which merge with the FB
at conical intersection point located at the edge of the Brillouin
zone \cite{flach14}. The tight-binding (discrete) model governing
the propagation of waves through this system is based on the
following equations:
\begin{eqnarray}
&&i\frac{da_{n}}{dt}+\alpha (b_{n}+b_{n+1})+\beta |a_{n}|^{2}a_{n}=0,  \notag
\\
&&i\frac{db_{n}}{dt}+\alpha (a_{n}+a_{n-1}+c_{n}+c_{n-1})+\beta
|b_{n}|^{2}b_{n}=0,  \label{first} \\
&&i\frac{dc_{n}}{dt}+\alpha (b_{n}+b_{n+1})+\beta |c_{n}|^{2}c_{n}=0,  \notag
\end{eqnarray}%
where $\alpha $ is the nearest-neighbor coupling strength and $\beta $ the
nonlinearity coefficient. These discrete Gross-Pitaevskii equations (GPEs)
describe a BEC trapped in the deep optical lattice. The same system can be
realized in optics, as an array of transversely coupled waveguides. In that
case, time $t$ is replaced by the propagation distance $z$. The evolution
equations (\ref{first}) can be derived from the Hamiltonian
\begin{gather}
\mathcal{H}=\sum_{n}\left\{ \epsilon _{a,n}|a_{n}|^{2}+\epsilon
_{b,n}|b_{n}|^{2}+\epsilon _{c,n}|c_{n}|^{2}+\frac{\beta }{2}%
(|a_{n}|^{4}\right. \\
\left. +|b_{n}|^{4}+|c_{n}|^{4})+\alpha \left[ (a_{n}^{\ast }+c_{n}^{\ast
})(b_{n}+b_{n+1})+\mathrm{c.c.}\right] \right\} ,
\end{gather}%
which is conserved, along with the norm,
$N=\sum_{n}(|a_{n}|^{2}+|b_{n}|^{2}+|c_{n}|^{2})$.

In the linear limit, $\beta =0$, the modal profiles, $\mathbf{\psi }%
_{n}=\{a_{n},b_{n},c_{n}\}$ are looked for as
\begin{equation}
\{{a}_{n}(t),{b}_{n}(t),{c}_{n}(t)\}=\mathbf{\Psi }_{n}e^{-iEt}.  \label{E}
\end{equation}
Using the Bloch basis, $\mathbf{\Psi }_{n}=\mathbf{\Psi }e^{ikn}$, with
wavenumber $k$ and the polarization eigenvectors
\begin{equation}
\mathbf{\Psi }^{(0)}=\frac{1}{\sqrt{2}}(1,0,-1),\mathbf{\Psi }^{(\pm )}=%
\frac{1}{2}\left( 1,\pm \frac{1+e^{ik}}{\sqrt{1+\cos k}},1\right) ,
\label{eqnosocmode}
\end{equation}
we obtain the band structure which consists of three branches
\begin{equation}
E_{\mathrm{FB}}(k)=0\;,\;E_{\pm }(k)=\pm 2\sqrt{2}\alpha \cos (k/2).
\label{spectrum}
\end{equation}
Two DBs are $k$-dependent branches, and the $k$-independent one is the FB,
see Fig.~\ref{fig:diamond}(b).

Henceforth, we set $\alpha =1$, by means of rescaling. The FB
states with $E=0$ in Eq. (\ref{eqnosocmode}) are independent of
$k$, and may be perfectly localized in a single cell in the form
of the linear CLS \cite{flach14},
\begin{equation}
\mathbf{\Psi }_{n}=\delta _{n,n_{0}}\mathbf{\Psi }^{(0)}\equiv \delta
_{n,n_{0}}\frac{1}{\sqrt{2}}(1,0,-1),  \label{nosoc}
\end{equation}%
where $\delta _{i,j}$ is the Kronecker's symbol, and $n_{0}$ determines the
location of the cell. Diffractive decay of the CLS is prevented by the
opposite signs of the field amplitudes at sites $a$ and $c$ and the
corresponding destructive interference.

The CLSs are, strictly speaking, infinitely degenerate states, as they can
be positioned in any unit cell. Any superposition of CLSs eigenvectors is
also an eigenvector. The present CLS belongs to class $U=1$, as it occupies
only one unit cell.

\subsection{The two-component system}

The basic model in our study is a two-component (pseudospinor)
system carried by the diamond chain, with the (pseudo-) SOC of
strength $\lambda $ which induces the linear mixing between the
two components, following the lines of Ref. \cite{discreteso}
(where the SOC of the Rashba type \cite{SOC} was adopted). The
components represent two hyperfine states of the same atomic
species in the binary BEC. One can think of the diamond chain as a
stripe embedded into a two-dimensional network, with the wave
functions vanishing at all other sites. The anisotropy of the SOC
defines a direction which forms some angle with the
diamond-chain's axis connecting all $b$ sites. Below we set this
angle to be $\pi /4$, although the cases of $0$ or $\pi /2$, as
well as the general case of an arbitrary angle can be readily
considered as well. Thus, the positive $x$-axis is directed from
site $b$ to $c$ in a unit cell, and the positive $y$-axis from $b$
to $a$ in the same cell. The two different components of the
discrete wave functions are denoted $a_{n}^{\pm },b_{n}^{\pm
},c_{n}^{\pm }$. We can also add a (pseudo) magnetic field $B$
which induces a Zeeman splitting, in terms of SOC. The
corresponding system of discrete GPEs is
\begin{eqnarray}
&&i\frac{da_{n}^{+}}{dt}+Ba_{n}^{+}+b_{n}^{+}+b_{n+1}^{+}+\lambda \left(
b_{n+1}^{-}+ib_{n}^{-}\right) +(\gamma |a_{n}^{+}|^{2}+\zeta
|a_{n}^{-}|^{2})a_{n}^{+}=0,  \notag \\
&&i\frac{db_{n}^{+}}{dt}%
+Bb_{n}^{+}+a_{n}^{+}+a_{n-1}^{+}+c_{n}^{+}+c_{n-1}^{+}+\lambda \left[
c_{n}^{-}-a_{n-1}^{-}-i(a_{n}^{-}-c_{n-1}^{-})\right] +(\gamma
|b_{n}^{+}|^{2}+\zeta |b_{n}^{-}|^{2})b_{n}^{+}=0,  \notag \\
&&i\frac{dc_{n}^{+}}{dt}+Bc_{n}^{+}+b_{n}^{+}+b_{n+1}^{+}-\lambda \left(
b_{n}^{-}+ib_{n+1}^{-}\right) +(\gamma |c_{n}^{+}|^{2}+\zeta
|c_{n}^{-}|^{2})c_{n}^{+}=0,  \notag \\
&&i\frac{da_{n}^{-}}{dt}-Ba_{n}^{-}+b_{n}^{-}+b_{n+1}^{-}-\lambda \left(
b_{n+1}^{+}-ib_{n}^{+}\right) +(\gamma _{1}|a_{n}^{-}|^{2}+\zeta
|a_{n}^{+}|^{2})a_{n}^{-}=0,  \notag \\
&&i\frac{db_{n}^{-}}{dt}-Bb_{n}+a_{n}^{-}+a_{n-1}^{-}+c_{n}^{-}+c_{n-1}^{-}-%
\lambda \left[ c_{n}^{+}-a_{n-1}^{+}+i(a_{n}^{+}-c_{n-1}^{+})\right]
+(\gamma _{1}|b_{n}^{-}|^{2}+\zeta |b_{n}^{+}|^{2})b_{n}^{-}=0,  \notag \\
&&i\frac{dc_{n}^{-}}{dt}-Bc_{n}^{-}+b_{n}^{-}+b_{n+1}^{-}+\lambda \left(
b_{n}^{+}-ib_{n+1}^{+}\right) +(\gamma _{1}|c_{n}^{-}|^{2}+\zeta
|c_{n}^{+}|^{2})c_{n}^{-}=0,  \label{socsveeq}
\end{eqnarray}
where two types of the local nonlinear terms are included:
collisions between atoms belonging to the same component generate
the cubic self-interaction with coefficients $\gamma $ and $\gamma
_{1}$ in Eq. (\ref{socsveeq}), while collisions between atoms from
different components give rise to the cross-interaction accounted
for by coefficient $\zeta $.

The above set of equations is invariant under the symmetry operation, which
involves the permutation of the SOC components, simultaneous sign change of
the magnetic field, the complex conjugation, and time reversal:
\begin{equation}
\{a,b,c\}^{\pm }\rightarrow \{a^{\ast },b^{\ast },c^{\ast }\}^{\mp
}\;,\;B\rightarrow -B\;,\;t\rightarrow -t\;.  \label{symmetry}
\end{equation}%
Note that in the nonlinear case this symmetry holds only if $\gamma =\gamma
_{1}=\zeta $.

\begin{figure}[h]
\center\includegraphics [width=10cm]{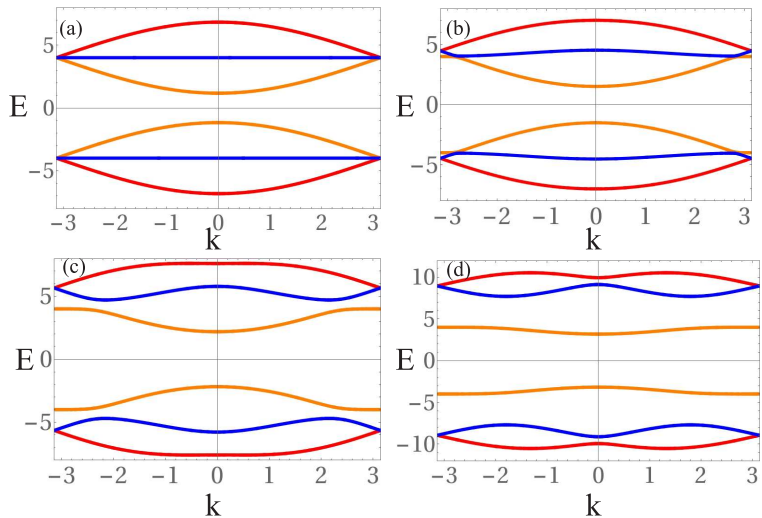}
\caption{Dispersion relations for the two-component system with
$B=4$ and
(a) $\protect\lambda =0$, (b) $\protect\lambda =1$, (c) $\protect\lambda =2$%
, (d) $\protect\lambda =4$. The Zeeman term $\sim B$ acts against the SOC
term $\sim \protect\lambda $, making the former FBs dispersive (curved).}
\label{fig3}
\end{figure}

The linear system with $\gamma =\gamma _{1}=\zeta =0$ and in the absence of
the SOC and magnetic fields, $\lambda =B=0$, is characterized by a
double-degenerate single-component spectrum of the diamond chain from the
previous section. Therefore, the FB exists and is double degenerate too, due
to the presence of two pseudospin components. For $\lambda =0$ but $B\neq 0$%
, the spectrum of each component is shifted by $\pm B$ relative to
the single-component model outlined in the previous section, see
Fig. \ref{fig3} (a). Thus, both FBs survive, but their degeneracy
is lifted. When the SOC is present, $\lambda \neq 0$, together
with the Zeeman terms, the FBs are deformed (made dispersive),
losing the flatness, see Fig. \ref{fig3}(b-d).

Being interested in effects related to the presence of the FB with
nonvanishing SOC, in what follows below we focus on the case when the SOC is
present ($\lambda \neq 0$), without the Zeeman splitting, $B=0$.

Besides the analytically found CLS solutions, other types of
stationary solutions of the nonlinear system are found below by
dint of a numerical procedure based on the Powell method
\cite{nashsoc}. The linear stability analysis of all found
localized solutions was performed numerically, solving linearized
equations for small perturbations which determine the stability
eigenvalues (EVs). The linearized equations are derived following
the well-known procedure \cite{nashsoc,nashnum}. Namely, small
perturbations are added to the localized solutions whose stability
is under consideration: $\mathbf{\tilde{\psi}}=\mathbf{\psi
}+\delta \mathbf{\psi }$, $|\delta \mathbf{\psi }|\ll
|\mathbf{\psi }|$, where $\mathbf{\tilde{\psi}}$ and $\mathbf{\psi
}$ denote, respectively, the multi-component discrete wave
function of the perturbed solution, and of its stationary
(unperturbed) counterpart. Perturbation eigenmodes are looked for
as $\delta \psi =\phi \exp (\rho t)$, where $\rho $ is the EV.
Generally, $\rho $ is a complex number whose positive real part,
if any, implies the exponential instability of the solution. Next,
the perturbed solution is substituted into Eq. (\ref{socsveeq}),
which is linearized with respect to the small perturbations. With
the help of a straightforward but cumbersome algebraic procedure,
the resulting linear system may be reduced to the EV problem of
the corresponding evolution matrix \cite{nashnum}. Finally, the
latter problem is solved in a numerical form. After that, results
predicted by the linear-stability analysis are verified against
direct simulations of Eq. (\ref{socsveeq}), using the Runge-Kutta
algorithm of the sixth order.

\section{Linear compact localized states in the two-component system with
the SOC}

\begin{figure}[h]
\center\includegraphics [width=10cm]{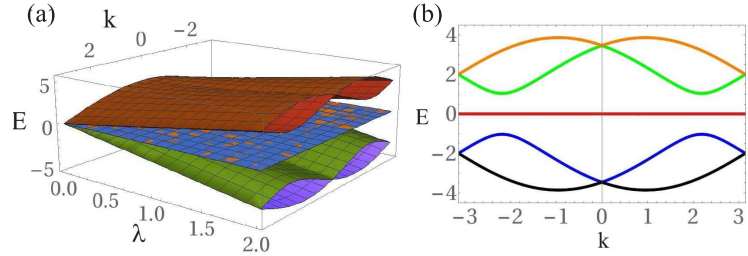} \caption{(a) The
effect of the SOC with strength $\protect\lambda $ on dispersion
relations (\protect\ref{bands}). (b) The dispersion relations
explicitly shown for $\protect\lambda =1$.} \label{fig1}
\end{figure}

For the case $B=0$ and $\lambda \neq 0$, the following branches of the
linear dispersion for eigenmodes taken as in Eq. (\ref{E}) can be obtained
from Eq. (\ref{socsveeq}):%
\begin{eqnarray}
&&E_{1,2}=0,  \notag \\
&&E_{3,4}=\pm 2\sqrt{1+\lambda ^{2}+\cos k-\sqrt{2}\lambda \left\vert \sin
k\right\vert },  \notag \\
&&E_{5,6}=\pm 2\sqrt{1+\lambda ^{2}+\cos k+\sqrt{2}\lambda \left\vert \sin
k\right\vert }.  \label{bands}
\end{eqnarray}%
These dispersion curves are plotted in Fig. \ref{fig1}, which
demonstrate that, in accordance with Eq. (\ref{bands}), the
double-degenerate FBs survive keeping their eigenvalues
$E_{1,2}=0$, unaffected by the SOC. This is by no means trivial,
since the SOC is lowering the system symmetry. Further, the FBs
are separated from the DBs by minigaps, whose widths,
$W_{\mathrm{mg}}$, increase with $\lambda $, as seen in Fig.
\ref{fig1}(a):
\begin{equation}
W_{\mathrm{mg}}(\lambda )=2\sqrt{1+\lambda ^{2}-\sqrt{1+2\lambda ^{2}}}.
\label{shirina}
\end{equation}%

The two mutually symmetric SIGs, with $E>0$ and $E<0$, are located,
respectively, above and beneath the DBs -- namely, at
\begin{equation}
|E|>E_{\min }^{\mathrm{(SIG)}}\equiv 2\sqrt{1+\lambda
^{2}+\sqrt{1+2\lambda ^{2}}}.  \label{SIG}
\end{equation}%
Values $E_{\min }^{\mathrm{(SIG)}}$ correspond to maximum values
of dispersion branches $E_{5,6}$ given by Eq. (\ref{bands}), which
are attained at $k=\pm \arctan \left( \sqrt{2}\lambda \right)$.

Two analytical solutions for the CLSs inside the FBs of the linear system ($%
\gamma =\gamma _{1}=\zeta =0$) are
\begin{align}
\mathbf{\Psi }_{n}^{(1)}=& C\{\delta _{n,n_{0}}\left( \frac{\lambda ^{2}-1}{2%
}(1+i),0,\frac{\lambda ^{2}+1}{2}-i\frac{\lambda ^{2}-1}{2},\lambda
(1-i),0,i\lambda \right)  \notag \\
& +\delta _{n+1,n_{0}}\left( \frac{\lambda ^{2}-1}{2}-i\frac{\lambda ^{2}+1}{%
2},0,\frac{\lambda ^{2}+1}{2}(1+i),\lambda ,0,0\right) \},  \notag \\
\mathbf{\Psi }_{n}^{(2)}=& C\{\delta _{n,n_{0}}\left( \frac{\lambda ^{2}-1}{2%
}-i\frac{\lambda ^{2}+1}{2},0,-\frac{\lambda ^{2}-1}{2}(1+i),-i\lambda
,0,\lambda (1+i)\right)  \notag \\
& +\delta _{n+1,n_{0}}\left( -\frac{\lambda ^{2}+1}{2}(1+i),0,\frac{\lambda
^{2}+1}{2}-i\frac{\lambda ^{2}-1}{2},0,0,\lambda \right) \},  \label{soclin}
\end{align}%
where $C$ is an arbitrary constant which determines the norm of
the wave function, i.e., the number of atoms, in terms of the BEC
model. Both solutions are CLSs with $U=2$, as they occupy two unit
cells. In the limit of $\lambda \rightarrow 0$, they split into
two obvious CLS states (\ref{nosoc}) set in adjacent cells. Thus,
the SOC induces a change of the CLS class from $U=1$ at $\lambda
=0$ to $U=2$ at $\lambda \neq 0$. The application of
transformation (\ref{symmetry}) produces two additional
symmetry-related solutions. The total norm of each of these
solutions is given by $|\Psi |^{2}=2|C|^{2}(\lambda ^{2}+1)^{2}$,
as follows from Eq. (\ref{soclin}).

Localized solutions are characterized by their
\textit{participation number}, which indicates the number of
strongly excited sites:
\begin{equation}
P_{t}=\left( \sum_{n}|\Psi _{n}|^{2}\right) ^{2}{\LARGE /}\sum_{n}|\Psi
_{n}|^{4},\quad \mathbf{\Psi }_{n}\equiv \left\{
a_{n}^{+},b_{n}^{+},c_{n}^{+},a_{n}^{-},b_{n}^{-},c_{n}^{-}\right\} ,
\label{part-ratio}
\end{equation}%
and by the spatial decay rate of the localized-mode's tail vs. $n$.

For the linear-CLS solutions (\ref{soclin}) obtained above, the
participation number is
\begin{equation}
P_{t}=\frac{4(\lambda ^{2}+1)^{4}}{(1/4)\left[ (\lambda ^{2}-1)^{4}+(\lambda
^{2}+1)^{4}\right] +(1/2)(\lambda ^{4}+1)^{2}+6\lambda ^{4}}\;,
\label{soclinpt}
\end{equation}
which yields the correct limit value $P_{t}(\lambda =0)=4$ in the
limit of vanishing SOC. In the limit of strong SOC, it returns to
the same limit value, $P_{t}(\lambda \rightarrow \infty )=4$
(practically, it is attained for $\lambda \geq 100$). At two
values of the SOC strength, $\lambda =\sqrt{2\mp \sqrt{3}}\approx
\left( 0.52,1.93\right)$, expression (\ref{soclinpt}) attains a
maximum, $P_{t}=6$. At an intermediate point, $\lambda =1$, there
is a local minimum, $P_{t}=16/3$.

The localized modes in the linear system are restricted to CLSs, sitting
solely on the FBs, with $E=0$, see Eq. (\ref{bands}). In the absence of the
SOC, there is no coupling between the equations for the two components,
hence CLSs are generated in each component independently, the respective
general solutions being linear superpositions of the simple stationary modes
given by Eq. (\ref{nosoc}).

\begin{figure}[h]\includegraphics [width=9cm]{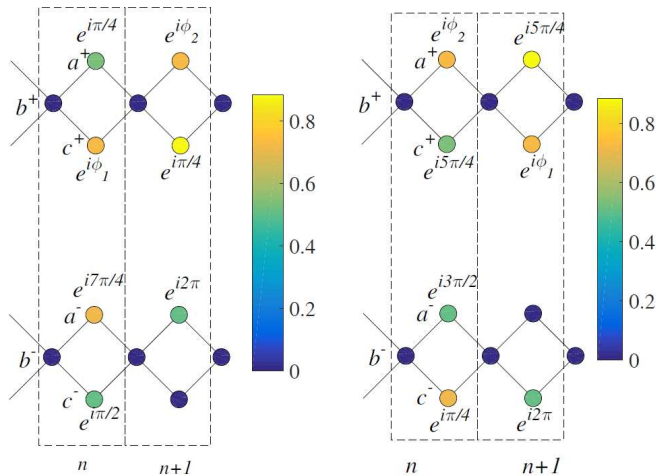}
\center\caption{FB-CLS modes in the linear system, $\Psi
_{n}^{(1)}$ (left) and $\Psi _{n}^{(2)}$ (right), as given by Eq.
(\protect\ref{soclin}). Values of the phase, marked in the figure
near the respective lattice sites, are $\protect\phi _{1}=-\arctan \left( \protect\lambda ^{2}-\left( \protect%
\lambda ^{2}-1\right) /\left( \protect\lambda ^{2}+1\right)
\right)$ and $\protect\phi _{2}=-\arctan \left( \left(
\protect\lambda ^{2}+1\right) /\left( \protect\lambda
^{2}-1\right) \right) $. The color code displays the squared
absolute value of the wave function on each site and for each
component. Note that the amplitudes strictly vanish on all
$b$-sites. Here, $\protect\lambda =2$.} \label{figls1}
\end{figure}

The SOC terms alter the shape of the CLS according to Eq. (\ref{soclin}). In
that case, the two components are coupled, and the structure of the compact
modes changes. They feature complex field amplitudes, and, as said above,
occupy (at least) two unit cells, see Fig. \ref{figls1}.

\section{Nonlinear localized modes}

\subsection{ Nonlinear CLSs and discrete solitons in the minigaps}

The next, obviously important, objective is to explore the impact of
nonlinearity in Eq. (\ref{socsveeq}). We find that the CLSs survive solely
if both the self- and cross-interactions are present, and the corresponding
nonlinearity parameters are \emph{exactly equal}, $\gamma =\gamma _{1}=\zeta
$. This is clearly related to the above-mentioned validity condition for
symmetry operation (\ref{symmetry}) in the nonlinear system. Being
interested in self-trapped modes, we consider the case of attractive
nonlinearity, with
\begin{equation}
\gamma =\gamma _{1}=\zeta >0.  \label{>0}
\end{equation}
For the opposite sign of the nonlinearity, essentially the same modes are
obtained, with the opposite sign of the frequency, $E>0$.

The CLSs have precisely the same structure as given by Eq. (\ref{soclin})
for the linear system, while the frequency of the nonlinear CLS families is
\begin{equation}
|E_{\mathrm{CLS}}|=\left( \gamma /2\right) |C|^{2}(\lambda ^{2}+1)^{2}
\label{detuning}
\end{equation}%
[recall that $C$ is the amplitude of solution (\ref{soclin})].
Thus, the nonlinear CLS modes are found in the \emph{exact
analytical form}, although they exist solely under the strict
condition (\ref{>0}). Equation (\ref{detuning}) demonstrates that,
varying $|C|^{2}$, one can tune the nonlinear CLS frequency to any
positive value across the whole minigap, through the dispersive
branches, and into the SIG. In particular, the existence of the
CLSs in the spectral areas occupied by the dispersive bands
implies that the CLSs may be identified as \textit{embedded
solitons} \cite{embedded}, i.e., ones which are embedded into the
bands of dispersive waves.

Straightforward algebraic manipulations show that, for the
nonlinear CLS modes, which can be created only if $\zeta =\gamma
=\gamma _{1}$, frequency $E$ and the total norm are related as
\begin{equation}
N=4|E|/\gamma .  \label{N}
\end{equation}
This relationship does not depend on $\lambda $ and is not valid for
discrete solitons (with exponentially decaying tails) considered below.

\begin{figure}[h]
\center\includegraphics [width=6cm]{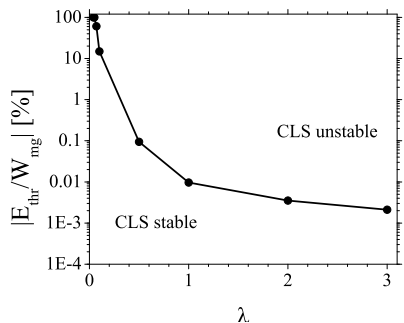} \caption{The
linear stability diagram for nonlinear CLSs. Stable CLSs exist for
$E<E_{\mathrm{thr}}$, which is plotted, in the log scale, in units
of the minigap width $W_{\mathrm{mg}}$ [see Eq.
(\protect\ref{shirina})].
Nonlinearity parameters are $\protect\gamma =\protect\gamma _{1}=\protect%
\zeta =1$.}
\label{lsacls}
\end{figure}

Both the linear stability analysis and direct simulations show
that the nonlinear CLSs are stable for frequencies $E$ staying in
a vicinity of the corresponding FB, \textit{viz}.,
$|E|<E_{\mathrm{thr}}$. According to Eqs. (\ref{detuning}) and
(\ref{N}), this implies that stable are CLSs with sufficiently
small amplitudes and norms. The dependence of the stability
threshold on the nonlinear-CLS frequency on $\lambda $ is shown in
Fig. \ref{lsacls} in units of the minigap's width,
$W_{\mathrm{mg}}$ (\ref{shirina}). In the limit of weak SOC,
almost the whole minigap is populated by stable nonlinear CLSs
(but note that the minigap itself becomes narrow in this case). On
the contrary, the increase of the SOC strength $\lambda $ leads to
a quick reduction of the relative stability threshold, which falls
to $<0.01$ for $\lambda >1$.

\begin{figure}[h]
\center\includegraphics [width=6cm]{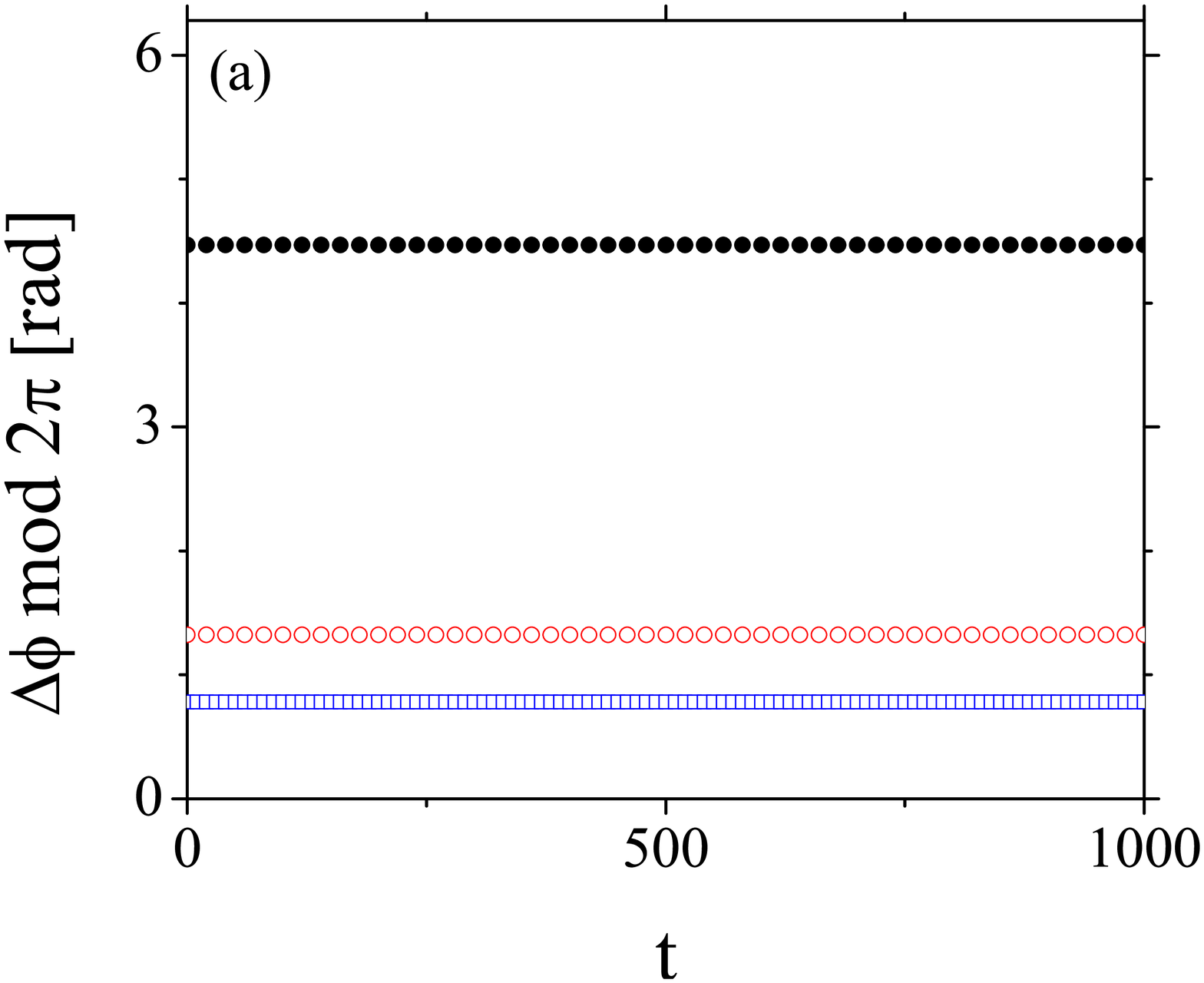}
\includegraphics
[width=6cm]{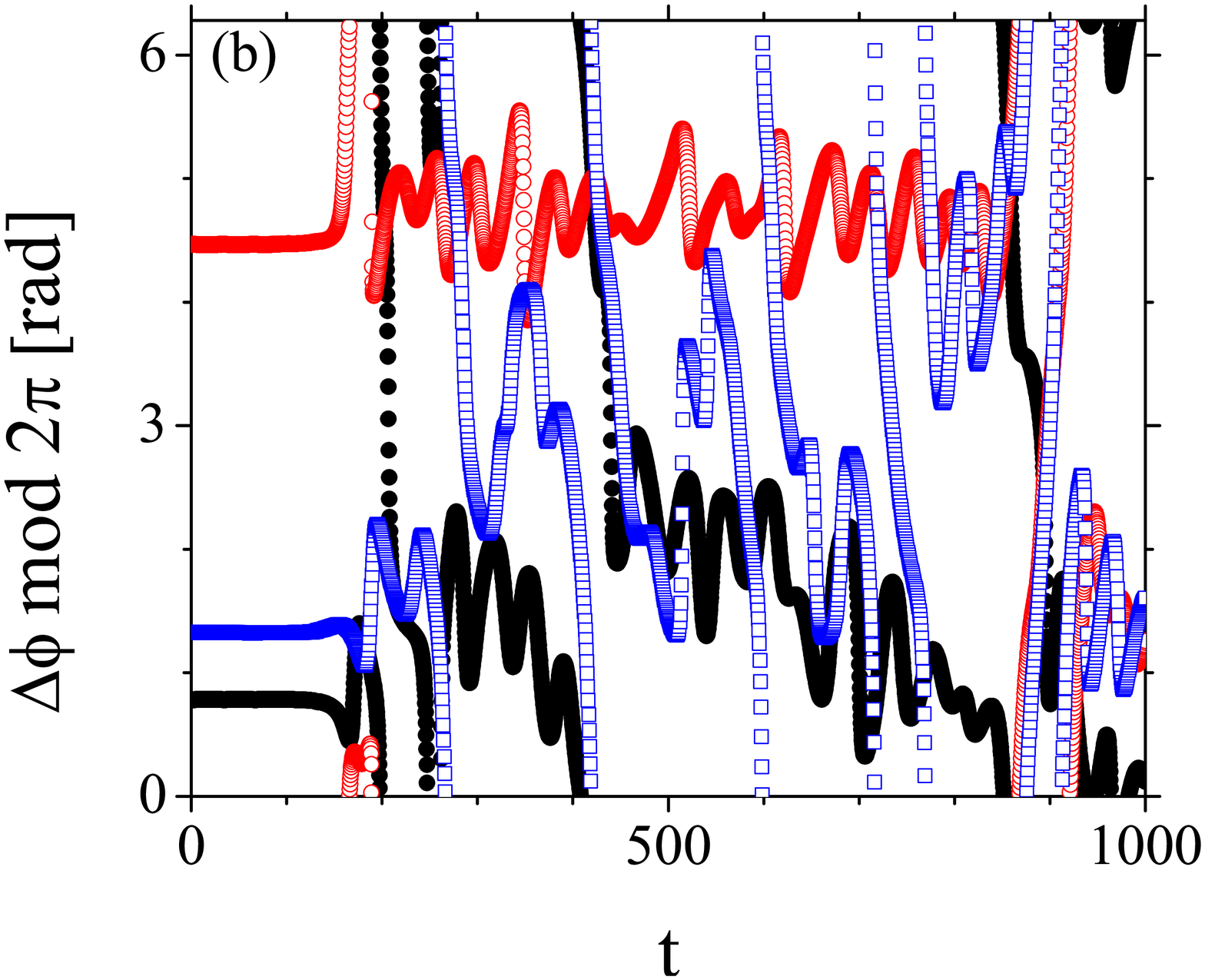} \caption{(a) The phase difference
between the field components of the stable (slightly perturbed)
CLSs at adjacent sites: $a_{n+1}^{+}$ and $a_{n}^{+}$ (black
circles), $c_{n+1}^{+}$ and $c_{n}^{+}$ (red empty circles),
$a_{n+1}^{-}$ and $a_{n}^{-}$ (blue squares), vs. time. (b) The
same for an unstable nonlinear CLSs. The SOC strength is
$\protect\lambda =2$. The
stable and unstable CLSs have, respectively, $E=-0.0009,\,N=0.0036$, and $%
E=-1.901,\,N=0.06$. The nonlinearity parameters are
$\protect\gamma =\protect\gamma _{1}=\protect\zeta =1$. }
\label{phasedif}
\end{figure}

\begin{figure}[h]
\center\includegraphics [width=10cm]{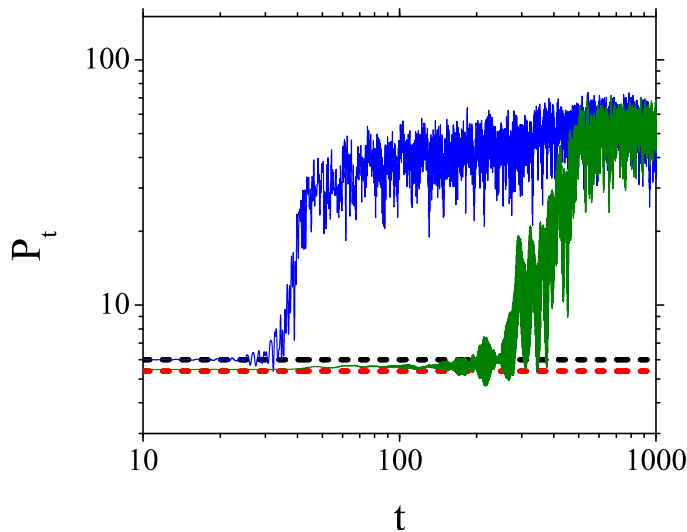} \caption{The
participation number of a stable nonlinear CLS, with
$E=-0.0009,\,N=0.0036$ (the black dashed line) and of an unstable
one, with $E=-1.901,\,N=0.06$ (the blue solid line) versus time.
Nonlinearity parameters are $\protect\gamma =\protect\gamma
_{1}=\protect\zeta =1$. Also shown are the time dependence of the
participation ratio for a stable ($E=-0.01,\,N=0.0036$, the red
dashed line) and unstable ($E=-0.49,\,N=0.27$, the green solid
line) discrete solitons (DSs). The SOC strength is
$\protect\lambda=2$. } \label{partcomp}
\end{figure}

To test the results of the linear-stability analysis, we simulated
perturbed evolution of the nonlinear CLSs in the framework of Eq.
(\ref{socsveeq}). To this end, the nonlinear CLSs, belonging to
the predicted stability and instability regions, with added a
small amplitude random perturbation \cite{nashsoc}, were used as
initial conditions. In the course of the evolution, stable
nonlinear CLSs keep constant phase differences between adjacent
lattice sites, see Fig. \ref{phasedif}. On the other hand,
unstable CLSs feature irregular variations of the phase
differences, which is accompanied by emission of energy into the
lattice background. The latter feature is illustrated in Fig.
\ref{partcomp} by plots for the participation ratio
(\ref{part-ratio}) of the respective modes.

\begin{figure}[h]
\center\includegraphics [width=14cm]{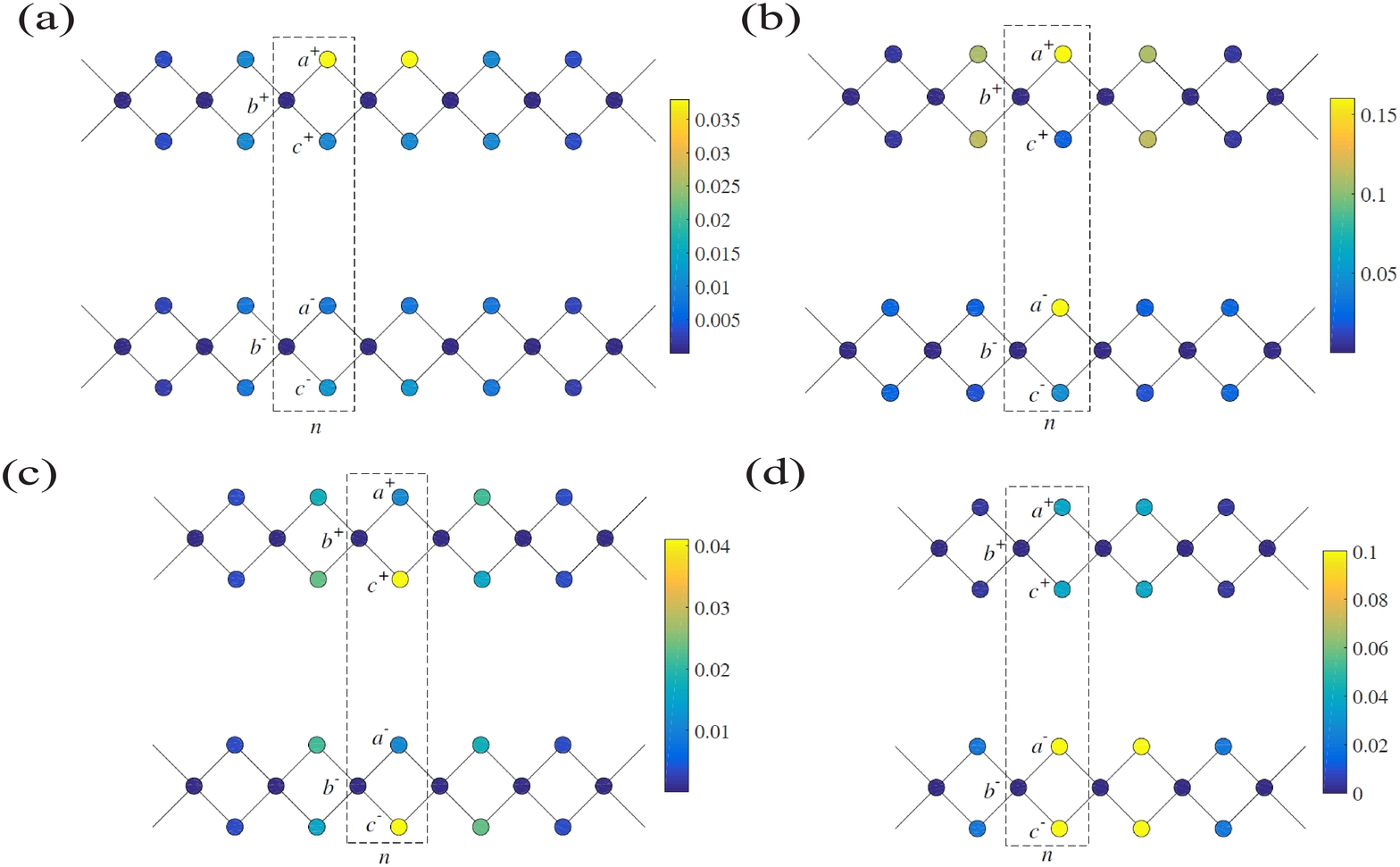}
\caption{Different discrete-soliton modes for $\protect\gamma
=\protect\gamma _{1}=1,\,\protect\zeta =0$ and $\protect\lambda
=2$.} \label{figmg0}
\end{figure}

\begin{figure}[h]
\center\includegraphics [width=14cm]{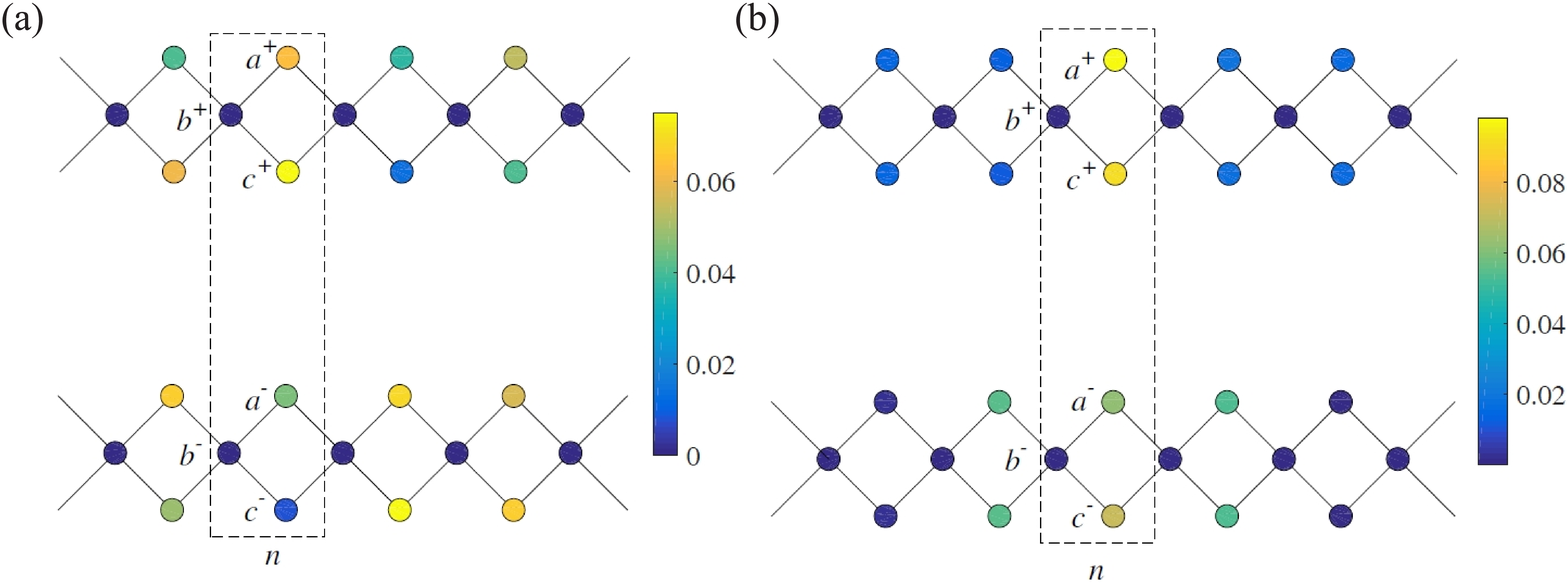}
\caption{Different discrete-soliton modes for $\protect\gamma
=\protect\gamma _{1}=\,\protect\zeta =1$ and $\protect\lambda
=2$.} \label{figmg1}
\end{figure}

\begin{figure}[h]
\center\includegraphics [width=8cm]{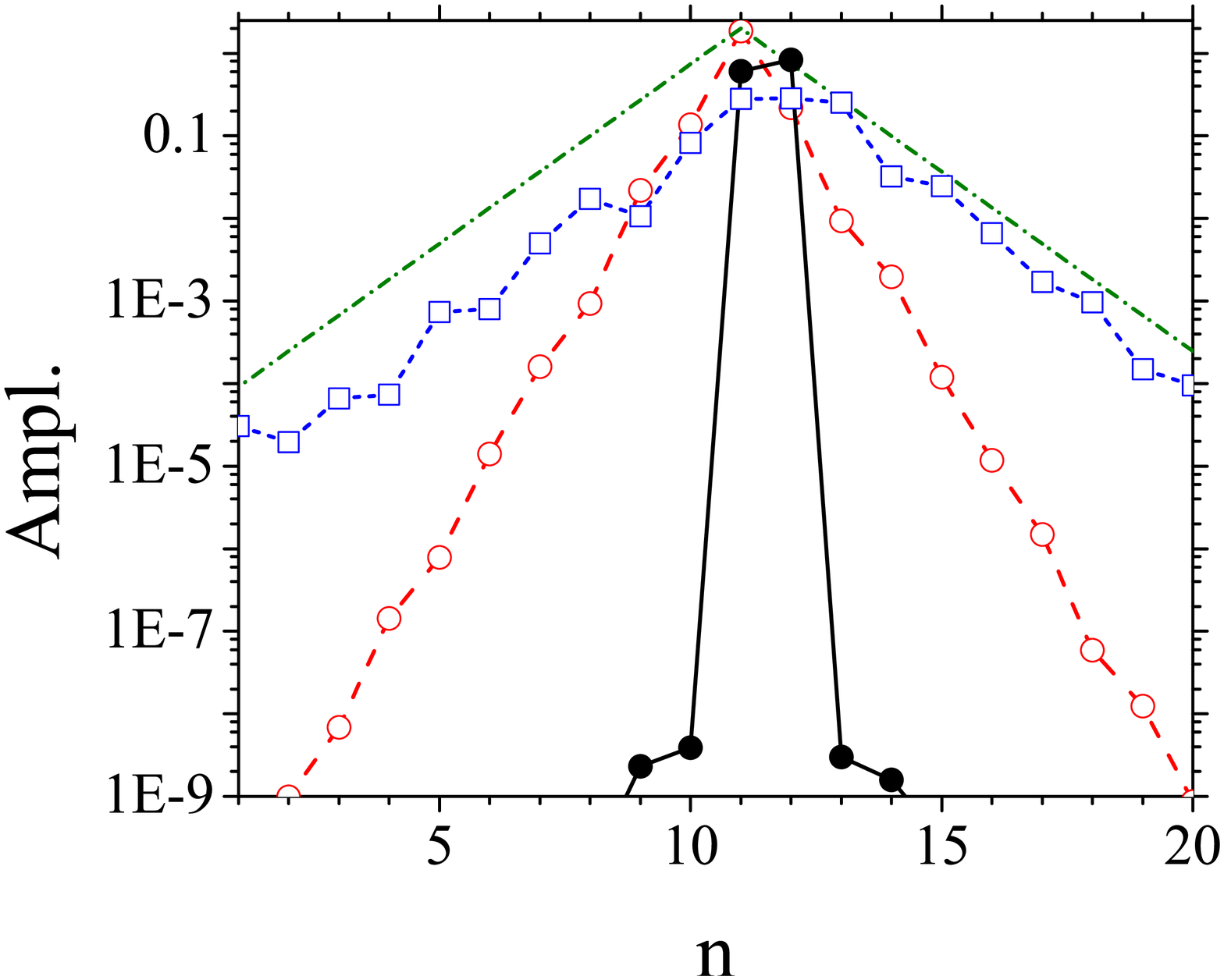} \caption{ The
spatial structure of normalized tails of different discrete
solitons on the log-log scale: a CLS (the black line with
circles), the discrete soliton, shown in Fig.
\protect\ref{figmg0}(d) (the blue dashed line with empty squares),
and the discrete soliton found in the semi-infinite gap (the red
line with empty circles). For comparison, the line with constant
slope $-1$ is shown as the olive dot-dashed one.
Parameters are $\protect\lambda =2$, $\protect\gamma _{1}=\protect\gamma =%
\protect\zeta =1$ for the modes denoted by solid lines, and $\protect\lambda %
=2$, $\protect\gamma _{1}=\protect\gamma =1$, $\protect\zeta =0$ for those
denoted by dashed lines. The values of $E$ and $N$ are the same as in Fig.
\protect\ref{partcomp}.}
\label{diverse}
\end{figure}

In addition to the nonlinear CLSs which persist in the linear
limit, the inclusion of the attractive nonlinearity gives rise to
exponentially localized discrete solitons (DSs), which are similar
to usual DSs in nonlinear lattices \cite{discrete}, see examples
in Figs. \ref{figmg0} and \ref{figmg1}. The structure of the DS
tails, and the comparison to the CLS shape [taken from Fig.
\ref{figmg0}(d)], are shown on the logarithmic scale in Fig.
\ref{diverse}. A drastic difference of the discrete solitons from
the CLSs is that the DSs exist as well when the cross-interaction
coefficient, $\zeta $, is different from its self-interaction
counterparts, $\gamma $ and $\gamma _{1}$ (for instance, in Fig.
\ref{figmg0} $\gamma =\gamma _{1}=1$ but $\zeta =0$).

Different DS modes are numerically obtained by changing the
respective input, i.e., its shape and norm, and ratios between the
self- and cross-interaction parameters, $\gamma $, $\gamma _{1}$,
and $\zeta $. Similar to the CLSs, the DSs have narrow stability
regions inside the minigaps. Exact location of boundaries of the
regions is a challenging issue, as they are sensitive to
variations of all the parameters. On the other hand, the existence
and stability of similar exponentially localized DSs in minigaps
was studied in other lattice models (see, e.g., Ref.
\cite{chaosmi}. Therefore, we here focus on accurate delineation
of the stability area for the CLSs in the minigaps of the
nonlinear lattice, which is a new problem.

The simulations demonstrate that perturbed stable states give rise
to robust breathers periodically oscillating in time. Outside of
the narrow stability regions found in the neighborhood of the FB,
unstable perturbed CLSs and DSs spread over the entire lattice in
the course of simulations. The evolution of participation number
for perturbed stable and unstable DSs of type shown in Fig.
\ref{figmg0} (d) is displayed in Fig. \ref{partcomp}, confirming
the predictions of the linear-stability analysis. The spreading
rate of unstable DSs is indeed determined by unstable EVs for
these modes.

Finally, it has been checked that stable CLSs and DSs may coexist
in a vicinity of the FB. They all are characterized by mutually
close (and small) values of the norm. In fact, their shapes are
close too, apart from the presence of the exponentially decaying
tails in the DSs, which are absent in the CLSs. The DS solutions'
existence area covers the entire minigaps, but their stability
region is a narrow strip adjacent to the FB. Unlike the CLSs, DS
states do not exists at values of $E$ belonging to the DBs, i.e.,
the DSs cannot be modes of the embedded type.

\subsection{Localized modes in the semi-infinite gap (SIG)}

\begin{figure}[h]
\center\includegraphics [width=10cm]{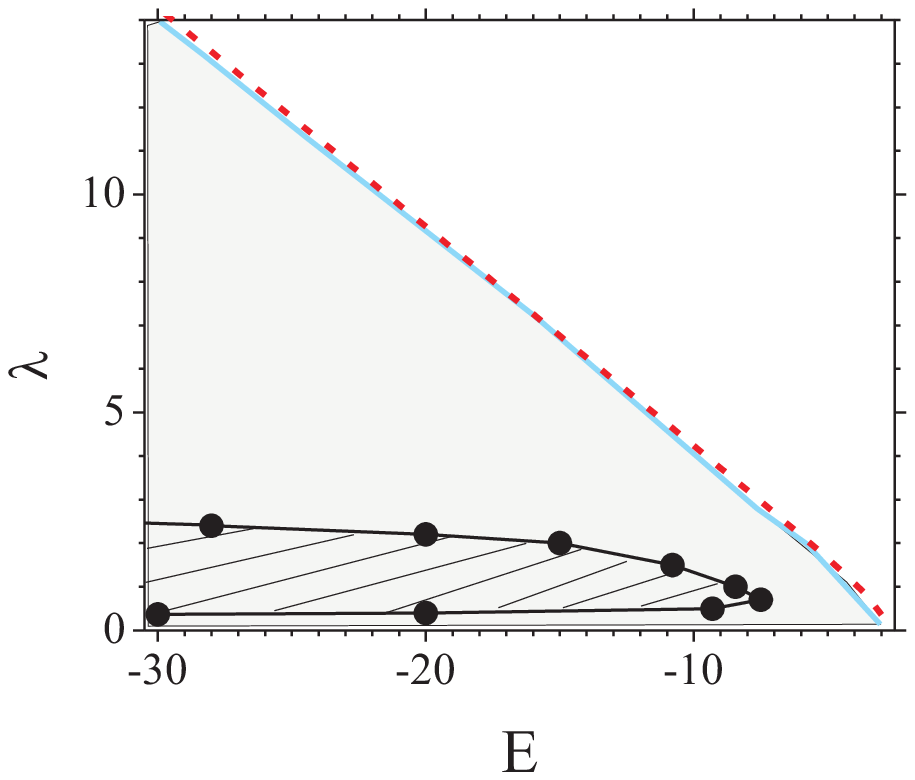} \caption{The
CLSs and one-peak DS found inside the SIG are stable in the shaded
and the gray area below the blue curve, respectively, in the plane
of $\left( E\text{,}\protect\lambda \right)$. The nonlinearity
parameters are fixed as $\protect\gamma =\protect\gamma
_{1}=\protect\zeta =1 $ for both solution types (recall that the
CLS does not exist if the these coefficients are not equal). The
dashed line shows the SIG boundary, as given by the inversion of
Eq. (\ref{SIG}): $\lambda =\sqrt{E^{2}/4+E/\sqrt{2}}$. }
\label{clsinsi}
\end{figure}

It is natural to explore the existence and stability of localized
states in the SIG. First, as mentioned above, the CLSs can be
found in the SIG. Note that the simple relation between the CLS
norm and frequency, given by Eq. (\ref{N}), applies to the SIG as
well. The stability area of the CLSs inside of SIG has been
identified too, through the linear-stability analysis and direct
simulations alike. The stability area in the plane of $\left(
E,\lambda \right) $ is shown in Fig. \ref{clsinsi}, cf. the
stability region for the CLSs in the minigaps, shown in Fig.
\ref{lsacls}. Similar to the fact that the CLSs are stable only a
small part of the minigap, the stability area occupies only a
small portion of the SIG.

\begin{figure}[h]
\center\includegraphics [width=14cm]{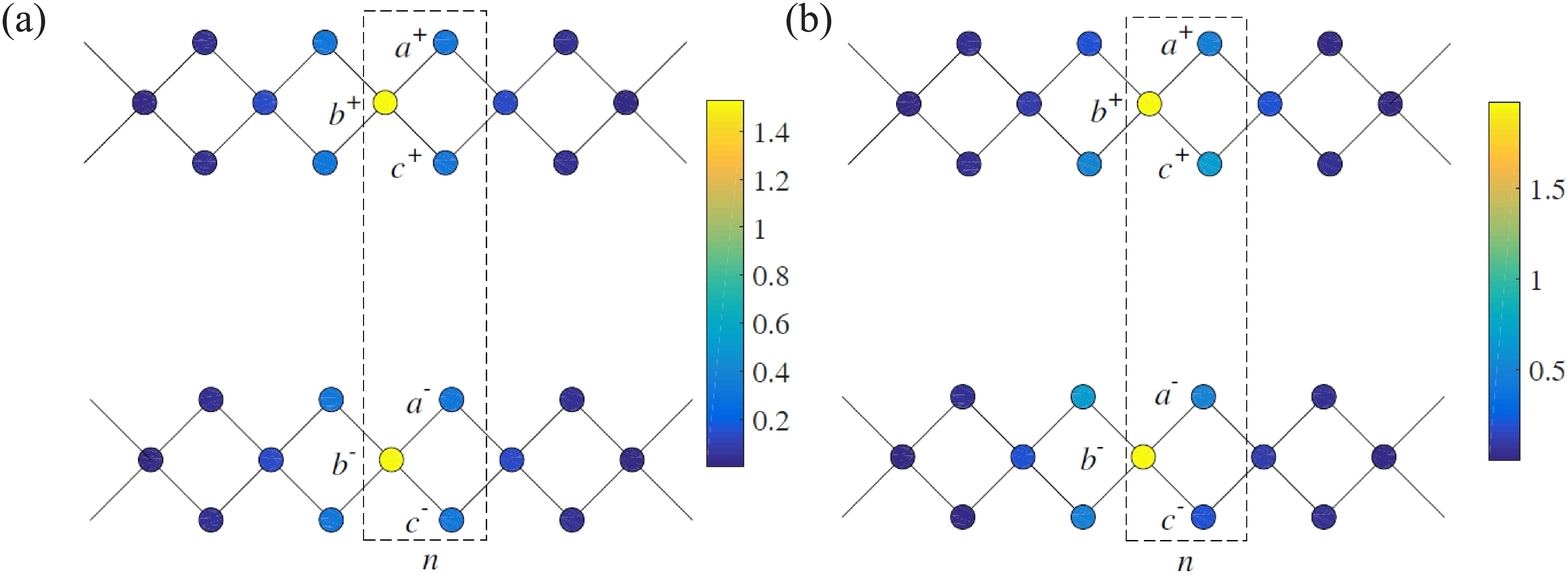}
\caption{Examples of stable DS mode found in the SIG, for
parameters (a) $\protect\lambda =0$, $E=-4.99,\,N=5.03$, and (b)
$\protect\lambda =1$, $E=-9.99,\,N=9.79$.} \label{fig13}
\end{figure}

In the absence of the SOC ($\lambda =0$), DSs, which are
characterized by the largest amplitude at site $b$ (on the
contrary to the CLSs, which have zero amplitudes at $b$), also
exist inside the SIG. They are found to be stable at $E<-5$ for
$\gamma ,\,\gamma _{1}>0$ (or at $E>5$ for $\gamma ,\,\gamma
_{1}<0$). A typical example of such a mode is displayed in Fig.
\ref{fig13}(a).

In the presence of the SOC, single-peak DSs are found in the SIG
too. Their symmetry with respect to the central $(b)$ site is
broken by the SOC. The stability region  of the single-peak DSs
almost overlaps with the SIG area, as shown in Fig. \ref{clsinsi}
(the gray area bounded by the blue solid curve). An example of the
stable DS is shown for $\lambda =2$, $\gamma =\gamma _{1}=\zeta
=1$ in Fig. \ref{fig13}(b). The affinity of these states to the
usual discrete solitons is confirmed by plotting tails of the
corresponding solutions in Fig. \ref{diverse} (the red line with
empty circles). For the sake of comparison, an exponentially
decaying function is shown by the green line in the same plot.

\section{Conclusions}

We have introduced a dynamical lattice system, which may be
realized for the two-component BEC with the SOC (spin-orbit
coupling), loaded into the optically imprinted ribbon with the
structure of the diamond chain, that serves as a paradigm for
realizing FB (flatband) spectra. The system includes interactions
between its components, both nonlinear and linear, the latter
mediated by the SOC. The linear version of the system, in the
absence of the SOC, is characterized by the spectrum which
consists of a pair of DBs (dispersive bands) which touch the FB at
the edge of the corresponding Brillouin zone, for each component.
Such a linear decoupled system creates CLS (compact localized
states) in each component which extend over one unit cell and
belong to the class $U=1$, at a single value of the frequency
which exactly corresponds to the FB. The introduction of the SOC
changes the spectrum, keeping the double-degenerate FBs and
opening narrow minigaps between them and the DBs. The CLSs remain
available in the exact analytical form. They increase their size
to $U=2$, and can be viewed as a bound state of two CLSs of class
$U=1$, with an appropriate phase and amplitude modulation,
depending on the strength of SOC. In the presence of the onsite
self-attractive cubic nonlinearity, we have shown that the CLSs
persist in the form of exact analytical solutions. Their frequency
can be tuned throughout the minigap, as well as into the SIG
(semi-infinite gap). Parallel to that, usual DSs (discrete
solitons) with exponentially decaying tails have been found too,
in a numerical form, in the minigap and SIG alike. Both the CLS
and DS modes are stable inside of the minigap in narrow stripes
abutting on the FB. Stability areas for the CLS and DS families
have been found in the SIG too, also being small in comparison
with the SIG size.

A main motivation for the interest in CLSs and FBs is due to the
macroscopic degeneracy they enforce. Such macroscopic degeneracies
are usually removed due to external perturbations, leading to new
eigenstates, whose properties are sensitively depending on the
type of perturbation added. In this work we show that spin-orbit
coupling with cold atoms in optical lattices acts as a
degeneracy-preserving perturbation. We also find that properly
tuned interactions destroy the notion of a band structure, yet
preserve the CLSs which simply detune their frequencies. A
detuning of these interactions will finally destroy the CLSs.

In addition to that, in this work we have demonstrated that the
presence of the FB strongly affects not only the CLSs but also
regular DSs. It is expected that these DSs will keep their
properties even if a perturbation of the system will lend the FB a
weak curvature.

As a development of the analysis, it may be interesting to analyze
a fully two-dimensional version of the system. In particular,
modes in the form of stable localized vortices have not been found
in the present system, both with and without the SOC terms. It is
plausible that vortex modes may exist in the fully two-dimensional
lattice.

\section*{Acknowledgments}

G.G., A.M., and Lj.H. acknowledge support from the Ministry of
Education and Science of Serbia (Project III45010). The work of
B.A.M. is supported, in part, by the joint program in physics
between the National Science Foundation (US) and Binational
Science Foundation (US-Israel), through grant No. 2015616. This
author appreciates hospitality of the Vin\v{c}a Institute of
Nuclear Sciences at University of Belgrade (Serbia).
This work was supported by the Institute for Basic Science, South Korea (Project Code IBS-R024-D1).

\end{document}